\documentclass[a4paper]{article}

\usepackage[english]{babel}
\usepackage{parallel,enumitem}
\usepackage[a4paper,top=3cm,bottom=2cm,left=3cm,right=3cm,marginparwidth=1.75cm]{geometry}

\usepackage{amsmath,amsthm,amssymb,amsfonts}
\usepackage{graphicx}
\usepackage[colorinlistoftodos]{todonotes}
\usepackage[colorlinks=true, allcolors=blue]{hyperref}
\usepackage{multicol}
\theoremstyle{definition}

\theoremstyle{plain}

\title{Cosmic Inflation From Fluctuating Baby-Skyrme Brane}

\author{Emir Syahreza Fadhilla$^{\sharp,\ddagger}$\footnote{Corresponding author}, Bobby Eka Gunara$^{\sharp}$,\\ Agus Suroso$^{\sharp}$, and Ardian Nata Atmaja$^{\ddagger}$\\ \\
$^{\sharp}$ \textit{\small Theoretical High Energy Physics Research Division, }\\
\textit{\small Faculty of Mathematics and Natural Sciences,}\\
\textit{\small Institut Teknologi Bandung}\\
\textit{\small Jl. Ganesha no. 10 Bandung, Indonesia, 40132}
\\ {\small and} \\
$^{\ddagger}$\textit{\small Research Center for Quantum Physics, National Research and Innovation Agency (BRIN),}\\
\textit{\small Kompleks PUSPIPTEK Serpong, Tangerang 15310, Indonesia}\\
\\
\small email: 30221012@mahasiswa.itb.ac.id, bobby@itb.ac.id, agussuroso@fi.itb.ac.id, ardi002@brin.go.id}

\date{\today}

\begin{document}
\maketitle

\begin{abstract}
In this work, we explore the inflationary dynamics induced by small fluctuations on the Skyrme brane, characterized by a time-dependent perturbative function \( \tilde{\phi} \). In the low-energy regime, the model successfully reproduces standard inflation, with a potential term dictated by the Skyrmion at the brane. Gravity localization is achieved at the brane, and the lowest energy scale is established at the asymptotic boundary. The model demonstrates the capability to emulate standard inflation dynamics, resembling \( \tilde{\phi}^4 \) potential characteristics under certain conditions. At higher energy levels, the behaviour of \( \tilde{\phi} \) is contingent upon the Skyrme term coupling constant \( \lambda \), influencing reheating phases. The wave-like nature of fluctuations allows for energy transfer, resulting in a possibly lower reheating temperature. We also discuss the prospect of \( \lambda \) changing sign during inflation, presenting a non-standard coupling dependent on the matter field.
\end{abstract}
\section{Introduction}
Gravity theories in higher dimensional bulk have been a thriving topic in high-energy physics in recent decades. It was started by the proposal from \cite{Kaluza:1921tu,Klein:1926tv} of five-dimensional spacetime to unify gravity and electromagnetism and is further supported by the fact that the six spatial dimensions of the eleven-dimensional supergravity can be compactified through the on Calabi-Yau threefold, resulting in an effectively five-dimensional model with compact extra dimension \cite{Horava:1995qa,Horava:1996ma,Witten:1996mz}. Such higher-dimensional theories provide more freedom for the choice of spacetime geometry and topology. One of the most well-known implications for the existence of extra dimensions is the variations of energy scale along the extra dimensions for gravity theory in (\(4+n\))-dimensional bulk, resulting in localization of matter in a four-dimensional brane with the lowest energy scale \cite{Arkani-Hamed:1998jmv,Antoniadis:1998ig,Randall:1999ee,Randall:1999vf,Arkani-Hamed:1998sfv}, which implies that the extra dimensions are not necessarily small in order to confine the standard model particles to the visible brane.
One of the main predictions that have been widely studied is the cosmological implication of the braneworld \cite{Csaki:1999jh,Binetruy:1999ut,Brax:2004xh,Cline:2006pcr}. The main feature of the braneworld cosmology is the correction term in the Friedmann equations that is proportional to the squared energy density of matter, \(\rho^2\), with a coupling constant that is inversely proportional to the brane tension, \(\tau\), implying that the term is negligible for low energy cases \(\rho\ll\tau\). The high-energy nature of this model is then used for cosmic inflation where a gravitating scalar on the bulk is introduced and it is found that the braneworld assumption allows steeper potential terms while maintaining the slow-roll condition \cite{Maartens:1999hf,Copeland:2000hn,Liddle:2003gw,Felipe:2004wh}. The scalar fields also support localization where it is known that the compactification scale depends on the minima of the potential term \cite{Goldberger:1999uk,Goldberger:1999un,Cline:2000xn}, leading to the solution of the fine-tuning problem in Randall-Sundrum model. Thus, a promising generalization of the braneworld cosmology would be the introduction of a sigma model on the bulk with an internal geometry that coincides with the scalar fields for flat target space. Recent findings also show that the inflationary model with multiple scalar fields is a good model for predicting the large scalar fluctuations and the production of primordial black holes at early time \cite{Sakharov:1993qh,Baumann:2014nda,Fumagalli:2020adf,Palma:2020ejf,Iacconi:2023slv}.

In this work, we consider a cosmological model from a braneworld with a non-linear sigma model on the bulk known as the Skyrme branes. The Skyrme brane is a gravitating Skyrme model where the scalar fields live in the higher dimensional bulk and possess the internal geometry of a hypersphere. It is similar to a gravitating sigma model with a modification that a non-standard kinetic term, known as the Skyrme term, is added to the Lagrangian \cite{Skyrme:1961vq,Skyrme:1962vh}. The first proposal of this type of theory is given by \cite{Blanco-Pillado:2008coy} where the authors propose a model with quadratic and quartic kinetic terms in the Skyrme Lagrangian, living in a seven-dimensional bulk and the Skyrme fields depends only on the extra dimensions, implying the fields are global in the three-dimensional visible brane. From here, the generalization to the Skyrme branes with \(N=4+n\) dimensional bulk is straightforward by using Manton's construction of generalized harmonic maps through invariants of the strain tensor defined as the product of the Jacobian of the fields\footnote{In this formalism, the field is seen as a diffeomorphism from \(n\) dimensional extra dimensions to a \(n\) dimensional sphere as the target space of the Skyrme fields} with its transpose \cite{manton1987,Gunara:2018lma,Gunara:2021jvq}. One of the interesting cases is the baby-Skyrme brane that lives in a \(N=6\) dimensional bulk proposed by \cite{Kodama:2008xm} where all possible kinetic terms are present since the two-dimensional target space only allows two strain tensor invariants. It is known that the static case of this theory provides localization of fermion fields and the emergence of three generations of fundamental fermion.  

The first inflating solution in the baby-Skyrme brane was proposed by \cite{Brihaye:2010nf} where it is shown that stable solutions exist for small perturbations around static Skyrmion profile. This result motivates us to propose a cosmological model with a dynamic Skyrmion profile that is generated by fluctuations around its static solution. The introduction of time-dependence in the Skyrme field is expected to contribute to the dynamics of the inflation on the brane, analogous to the contribution of bulk scalar in the usual braneworld model, with a potential term that depends directly on the geometry of the target space of the field. 

\section{Baby-Skyrme Brane}
The baby-Skyrme brane, introduced in \cite{Blanco-Pillado:2008coy,Brihaye:2010nf} is an Einstein-Skyrme model in six dimensional bulk, \(\mathcal{B}\), where the action is given by
\begin{eqnarray}\label{Action}
    S&=&-\int\sqrt{-g}\left[C_0V(\phi)+C_1\frac{\partial \phi^a}{\partial X^\mu}\frac{\partial \phi^a}{\partial X^\nu}g^{\mu\nu}+\frac{C_2}{4}\frac{\partial \phi^a}{\partial X^{[\mu}}\frac{\partial \phi^b}{\partial X^{\alpha]}}\frac{\partial \phi^a}{\partial X^{[\nu}}\frac{\partial \phi^b}{\partial X^{\beta]}}g^{\mu\nu}g^{\alpha\beta}\right]d^6X\nonumber\\
    &&+\int\sqrt{-g}\frac{R-2\Lambda}{16\pi \mathcal{G}}d^6X,
\end{eqnarray}
and the bulk can be decomposed into \(\mathcal{M}\times \mathbb{R}^2\), \(\mathcal{M}\) is the four-dimensional spacetime, the metric on the bulk is given by the ansatz
\begin{equation}
    ds^2=F(z)\left(\gamma_{ij}dx^i dx^j\right)+G(z)\left(dz^2+z^2d\zeta^2\right),
\end{equation}
\(z\geq0\), \(0\leq\zeta<2\pi\) , and the metric on \(\mathcal{M}\) is the Friedmann–Lemaître–Robertson–Walker,
\begin{equation}
    \gamma_{ij}dx^i dx^j=-dt^2+a^2(t) \left(\frac{dr^2}{1-kr^2}+r^2d\Omega^2\right),
\end{equation}
where \(k\in\{-1,0,1\}\) depending on the curvature of the spatial submanifold of \(\mathcal{M}\). The Skyrme field, \(\phi\), is given by the multiplet \(\phi=(\phi^0,\phi^1,\phi^2)\) where \(\phi^a\phi^a=1\) must be assumed since the baby-Skyrme model obeys the \(O(3)\) sigma model constraint. Both \(C_1\) and \(C_2\) are the coupling constants of the kinetic term and the Skyrme term, respectively, and this model contains all possible non-standard kinetic terms for a Skyrme model that have a two-dimensional base space because the Skyrme field is assumed to be global in \(\mathcal{M}\). here, we assume that the ansatz is given by the time-dependent hedgehog ansatz where the profile of the Skyrme field depends only on the extra dimensions, namely
\begin{equation}
    \phi=(\cos\xi(t,z),\sin\xi(t,z)\cos(n\zeta),\sin\xi(t,z)\sin(n\zeta)),
\end{equation}
\(n\in\mathbb{Z}\) that is necessary in order to guarantee that the value of \(\phi\) is unique on all points in \(\mathcal{B}\).

From the action above we can derive the dynamical equations of the Skyrme field through variations with respect to \(\phi^a\) which gives
\begin{eqnarray}
                \left[\nabla^\mu\left[2C_1\phi^b_\mu+ 2{C_2}g^{\alpha\beta}\left(\phi^b_{\mu}\phi^a_\alpha\phi^a_\beta-\phi^b_\alpha\phi^a_\beta\phi^a_{\mu}\right)\right]-\frac{\partial}{\partial \phi^b} C_0 V\right]\left(\delta^{db}-\phi^d\phi^b\right)=0~,
            \end{eqnarray}
where we have denoted the derivative \(\partial\phi^a/\partial X^\mu\) as \(\phi^a_\mu\). On the other hand, the dynamical equations for the metric components from action \eqref{Action}, are given by Einstein's field equations
\begin{equation}
    G_{\mu\nu}+g_{\mu\nu}\Lambda=8\pi\mathcal{G}T_{\mu\nu},
\end{equation}
where \(T_{\mu\nu}=T^{\text{Skyrme}}_{\mu\nu}\), explicitly given by
\begin{eqnarray}
    T^{\text{Skyrme}}_{\mu\nu}&=&-g_{\mu\nu}\left[C_0V(\phi)+C_1\phi^a_\mu\phi^a_\nu g^{\mu\nu}+\frac{C_2}{4}\phi^a_{[\mu}\phi^b_{\alpha]}\phi^a_{[\nu}\phi^b_{\beta]}g^{\mu\nu}g^{\alpha\beta}\right]\nonumber\\
    &&+\left[2C_1\phi^a_\mu\phi^a_\nu+C_2\phi^a_{[\mu}\phi^b_{\alpha]}\phi^a_{[\nu}\phi^b_{\beta]}g^{\alpha\beta}\right].
\end{eqnarray}
We should make the corresponding equations dimensionless by introducing the transformation \((t,z)\rightarrow(t/\sqrt{C_1},r/\sqrt{C_1})\) and we can simplify the coupling constants by introducing \(\alpha=8\pi \mathcal{G}C_1\), \(\lambda=C_2\), and \(\sigma=C_0/C_1^2\).
As such, the system of equations that is going to be solved is given by
\begin{eqnarray}
    &&\frac{3 \left(\dot{a}^2+k\right)}{a^2}-\frac{3\left(F'+z
   F''\right)}{2zG}-\frac{F\left(G'+z G''\right)}{2zG^2}+\frac{ F G'^2}{2  G^3}-F\Lambda\nonumber\\&=&\alpha F\left[\left(\frac{1}{F}\dot{\xi}^2+\frac{1}{G}\xi'{}^2+n\frac{\sin^2\xi}{Gz^2}\right)+\lambda\left(\frac{1}{FG}\dot{\xi}^2\xi'{}^2+n\dot{\xi}^2\frac{\sin^2\xi}{FGz^2}+n\xi'{}^2\frac{\sin^2\xi}{G^2z^2}\right)+\sigma V\right],
\end{eqnarray}
from the \(tt\) component of Einstein's equation,
\begin{eqnarray}
    &&\frac{3  \left(\dot{a}^2+a \ddot{a}+k\right)}{a^2}-\frac{
   F' G'}{G^2}-\frac{3 F'^2}{2FG}-\frac{2 F'}{zG}-F\Lambda\\&=&\alpha F\left[\left(\frac{1}{F}\dot{\xi}^2+\frac{1}{G}\xi'{}^2-n\frac{\sin^2\xi}{Gz^2}\right)+\lambda\left(-\frac{1}{FG}\dot{\xi}^2\xi'{}^2+n\dot{\xi}^2\frac{\sin^2\xi}{FGz^2}+n\xi'{}^2\frac{\sin^2\xi}{G^2z^2}\right)-\sigma V\right]\nonumber,
\end{eqnarray}
from the \(zz\) component of Einstein's equation, 
\begin{eqnarray}
    &&\frac{3   \left(\dot{a}^2+\ddot{a}+k\right)}{a^2 }+\frac{ F' G'}{ G^2}-\frac{ F'^2}{2
   FG}-\frac{2  F''}{G}-F\Lambda\\&=&\alpha F\left[\left(\frac{1}{F}\dot{\xi}^2-\frac{1}{G}\xi'{}^2+n\frac{\sin^2\xi}{Gz^2}\right)+\lambda\left(\frac{1}{FG}\dot{\xi}^2\xi'{}^2-n\dot{\xi}^2\frac{\sin^2\xi}{FGz^2}+n\xi'{}^2\frac{\sin^2\xi}{G^2z^2}\right)-\sigma V\right]\nonumber,
\end{eqnarray}
from the \(\zeta\zeta\) component of the Einstein's equation, and
\begin{eqnarray}
    &&\frac{\dot{a}^2+2a\ddot{a}+k}{a^2}-\frac{3 \left(F'+zF''\right)}{2 z G
   }+\frac{ FG'^2}{2 G^3 }-\frac{ F \left(G'+zG''\right)}{2 z G^2 }-F\Lambda\\&=&\alpha F\left[\left(-\frac{1}{F}\dot{\xi}^2+\frac{1}{G}\xi'{}^2+n\frac{\sin^2\xi}{Gz^2}\right)+\lambda\left(-\frac{1}{FG}\dot{\xi}^2\xi'{}^2-n\dot{\xi}^2\frac{\sin^2\xi}{FGz^2}+n\xi'{}^2\frac{\sin^2\xi}{G^2z^2}\right)+\sigma V\right]\nonumber,
\end{eqnarray}
from the \(rr\) component of Einstein's equation while the rest of the components are equivalent to the \(rr\) component. From the Skyrme field dynamical equation, we have
\begin{eqnarray}
   &&-\frac{1}{a^3F} \frac{\partial}{\partial t}\left(a^3\Dot{\xi}\left[1+\lambda\left(\frac{\xi'{}^2}{G}+n\frac{\sin^2\xi}{Gz^2}\right)\right]\right)+\frac{1}{F^2Gz}\frac{\partial}{\partial z}\left(F^2z\xi'\left[1+\lambda\left(n\frac{\sin^2\xi}{Gz^2}-\frac{\dot{\xi}^2}{F}\right)\right]\right)\nonumber\\&=&\frac{n}{Gz^2}\sin\xi\cos\xi\left[1+\lambda\left(\frac{\xi'{}^2}{G}-\frac{\dot{\xi}^2}{F}\right)\right]+\frac{\sigma}{2}\frac{dV}{d\xi}.
\end{eqnarray}
From this point, we would like to focus ourselves on the case of pion-mass type potential, \(V=1-\phi^0=1-\cos\xi\) that has two vacua since \(0\leq\xi<\pi\), namely \(\xi=0\) and \(\xi\rightarrow\pi\).
Our main goal here is to find the solutions of four functions, namely \(a(t)\), \(\xi(t,z)\), \(F(z)\) and \(G(z)\), with three parameters \(n\), \(\lambda\) and \(\sigma\).
\section{Correspondence to The Standard Inflation Model}\label{correspondence}
The standard inflation model with Klein-Gordon scalar assumes \(k=0\) which implies that the spacetime is spatially flat and we expect that such a model is the low energy limit of our model since it is the relevant model before the particle creation at the end of the reheating phase. At the low energy limit, this model coincides with the sigma model with no Skyrme term because the Skyrme term is negligible on the large-length scale, i.e. \(\lambda\approx 0\). The assumptions mentioned above give us the following set of equations for the low energy limit,
\begin{eqnarray}
    3H^2-\frac{3\left(F'+z
   F''\right)}{2zG}-\frac{F\left(G'+z G''\right)}{2zG^2}+\frac{ F G'^2}{2  G^3}-F\Lambda&=&\alpha F\left[\left(\frac{1}{F}\dot{\xi}^2+\frac{1}{G}\xi'{}^2+n\frac{\sin^2\xi}{Gz^2}\right)+\sigma V\right],\nonumber\\
   3\dot{H}+6H^2-\frac{
   F' G'}{G^2}-\frac{3 F'^2}{2FG}-\frac{2 F'}{zG}-F\Lambda&=&\alpha F\left[\left(\frac{1}{F}\dot{\xi}^2+\frac{1}{G}\xi'{}^2-n\frac{\sin^2\xi}{Gz^2}\right)-\sigma V\right]\nonumber,\\
   3\dot{H}+6H^2+\frac{ F' G'}{ G^2}-\frac{ F'^2}{2
   FG}-\frac{2  F''}{G}-F\Lambda&=&\alpha F\left[\left(\frac{1}{F}\dot{\xi}^2-\frac{1}{G}\xi'{}^2+n\frac{\sin^2\xi}{Gz^2}\right)-\sigma V\right]\nonumber,\\
   2\dot{H}+3H^2-\frac{3 \left(F'+zF''\right)}{2 z G
   }+\frac{ FG'^2}{2 G^3 }-\frac{ F \left(G'+zG''\right)}{2 z G^2 }-F\Lambda&=&\alpha F\left[\left(-\frac{1}{F}\dot{\xi}^2+\frac{1}{G}\xi'{}^2+n\frac{\sin^2\xi}{Gz^2}\right)+\sigma V\right]\nonumber,\\
   -3\frac{H}{F}\dot{\xi}-\frac{\ddot{\xi}}{F}+\frac{1}{F^2Gz}\frac{\partial}{\partial z}\left(F^2z\xi'\right)&=&\frac{n}{Gz^2}\sin\xi\cos\xi+\frac{\sigma}{2}\frac{dV}{d\xi},\label{inflatonCorr}
\end{eqnarray}
with \(H=\dot{a}/a\). Let us Consider the last equation in the system \eqref{inflatonCorr}, with perturbative time-dependencies such that \(\xi(t,z)=\psi(z)+\Tilde{\phi}(t)\), and \(|\phi(t)/\left<\psi(z)\right>|<1\) for all \(t\). This way, we can expand 
\begin{eqnarray}
    \frac{dV}{d\xi}&=&\sin\xi=\sin\psi+\cos\psi\Tilde{\phi}-\frac{\sin\psi}{2!}\Tilde{\phi^2}-\frac{\cos\psi}{3!}\Tilde{\phi^3}+\dots,\nonumber\\
    \cos\xi&=&\cos\psi-\sin\psi\Tilde{\phi}-\frac{\cos\psi}{2!}\Tilde{\phi}^2+\frac{\sin\psi}{3!}\Tilde{\phi}^3+\dots,
\end{eqnarray}
such that the dynamical equation for \(\xi\) becomes
\begin{eqnarray}\label{perturbPhi}
    &&-3H\dot{\Tilde{\phi}}-\Ddot{\Tilde{\phi}}+\frac{1}{FGz}\left(F^2z\psi'\right)\nonumber\\&=&\frac{n}{Gz^2}\left[\sin\psi\left(\frac{\sigma Gz^2}{2n}+\cos\psi\right)+\left(\cos(2\psi)+\frac{\sigma Gz^2}{2n}\cos\psi\right)\Tilde{\phi}\right.\nonumber\\&&\left.-2\left(\sin\psi\cos\psi+\frac{\sigma Gz^2}{8n}\sin\psi\right)\Tilde{\phi}^2+\frac{1}{6}\left(-4\cos(2\psi)-\frac{\sigma Gz^2}{2n}\cos\psi\right)\Tilde{\phi}^3\right].
\end{eqnarray}
Let us assume that \(\psi\) is the unperturbed Skyrmion profile, then the lowest order of perturbations above is the effective field equations for the Skyrme field, namely
\begin{equation}
    \frac{1}{FGz}\left(F^2z\psi'\right)'=\sin\psi\left(\frac{\sigma }{2}+\frac{n}{Gz^2}\cos\psi\right),
\end{equation}
which should be solved for \(\psi\), together with the Einstein's equations. The remaining time-dependent part of the field equation is given by,
\begin{equation}
    0=3H\dot{\Tilde{\phi}}+\Ddot{\Tilde{\phi}}+m_1\Tilde{\phi}+m_2\Tilde{\phi}^2+m_3\Tilde{\phi}^3+\dots,
\end{equation}
where \(m_i\) are the coefficients given in \eqref{perturbPhi} that depends on the \(\psi,~F\) and \(G\) evaluated at \(z_0\) that is the location of the visible brane. We can see that the dynamical equation for \(\Tilde{\phi}\) is the same as the equation for the inflation from the Einstein-Klein-Gordon theory but the potential is deduced from the profile of the gravitating Skyrmion in the bulk.

The location of the visible brane is dictated by the energy scale that depends on the profile of \(F\) and \(G\) in the bulk. As such, we shall proceed by solving the unperturbed Einstein's equations that are given by,
\begin{eqnarray}\label{sigmaEInst1}
    -\frac{3\left(F'+z
   F''\right)}{2zF}-\frac{\left(G'+z G''\right)}{2zG}+\frac{ G'^2}{2  G^2}-G\Lambda&=&\alpha \left[\psi'{}^2+n\frac{\sin^2\psi}{z^2}+\sigma G (1-\cos\psi)\right],\\\label{sigmaEInst2}
   -\frac{
   F' G'}{FG}-\frac{3 F'^2}{2F^2}-\frac{2 F'}{zF}-G\Lambda&=&\alpha \left[\psi'{}^2-n\frac{\sin^2\psi}{z^2}-\sigma G (1-\cos\psi)\right],\\
   -\frac{ F' G'}{ FG}+\frac{ F'^2}{2
   F^2}+\frac{2  F''}{F}+G\Lambda&=&\alpha \left[\psi'{}^2-n\frac{\sin^2\psi}{z^2}+\sigma G (1-\cos\psi)\right].\label{sigmaEInst3}
\end{eqnarray}
From here, we are going to assume that the two branes are located at \(z=0\) and \(z\rightarrow\infty\), and the corresponding boundary conditions for the Skyrme branes are 
\begin{equation}
    \psi(0)=0,~~~\lim_{z\rightarrow\infty}\psi=\pi,
\end{equation}
such that the topological charge of the Skyrmion is equal to \(n\) and the stable vacuum is located at \(z\rightarrow\infty\). To proceed, let us expand \(\psi\), \(F\) and \(G\) near \(z=0\) as follows
\begin{eqnarray}
    \psi&=&\psi_1z+\psi_2z^2+O(z^3),\nonumber\\
    \label{expandzero}F&=&F_0+F_1z+F_2z^2+O(z^3),\\
    G&=&G_0+G_1z+G_2z^2+O(z^3)\nonumber.
\end{eqnarray}
The behaviour of each function near the asymptotic boundary is assumed to be frozen to constant values \cite{Gunara:2021jvq},
\begin{eqnarray}
    \psi&=&\pi+\frac{\bar{\psi}_1}{z^{p_1}}+O(z^{-(p_1+1)}),\nonumber\\ \label{expandinfty}    F&=&1+\frac{\bar{F}_1}{z^{p_2}}+O(z^{-(p_2+1)}),\\
   G&=&1+\frac{\bar{G}_1}{z^{p_3}}+O(z^{-(p_3+1)})\nonumber,
\end{eqnarray}
with all \(\psi_i,~F_i,~G_i,~\bar{\psi}_i,~\bar{F}_i,\) and \(\bar{G}_i\) are constants. This expansion can be used to find the behaviour of the fields near \(z=0\), by using the Einstein equation (\ref{sigmaEInst1}-\ref{sigmaEInst3}) up to the second order of \(z\), which leads to
\begin{eqnarray}
    n&=&1,\\
    F_1&=&0,\\
    F_2&=&-\frac{F_0G_0\Lambda}{4},\\
    G_1&=&0,\\
    G_2&=&\frac{G_0^2\Lambda}{4}.
\end{eqnarray}
This result implies that the topological charge of our setup must be equal to one, the value of function \(F\) is decreasing near \(z=0\) since it is a local maximum of \(F\), and the value of \(G\) is increasing since \(z=0\) is the local minima of \(G\) if \(\Lambda>0\). Another implication is that both \(F_0\) and \(G_0\) are shooting parameters whose values should be tuned in order to satisfy the boundary conditions at \(z\rightarrow\infty\).

Since we know that \(F\) is decreasing and \(G\) is increasing, given that the condition for \(\psi_1\) above is satisfied, then the energy scale at \(z\rightarrow\infty\) is lower than the scale at \(z=0\), which implies that the visible brane should be located at the asymptotic boundary of the bulk. As such, the constants \(m_i\)s should be evaluated at \(z\rightarrow \infty\). This result gives us, up to third-order perturbation, the following constants
\begin{eqnarray}
    m_1&=&-\frac{\sigma}{2}\nonumber,\\
    m_2&=&0\nonumber,\\
    m_3&=&\frac{\sigma}{12},
\end{eqnarray}
which leads to the following dynamics of \(\Tilde{\phi}\),
\begin{equation}
0=3H\dot{\Tilde{\phi}}+\Ddot{\Tilde{\phi}}-\frac{\sigma}{2}\Tilde{\phi}+\frac{\sigma}{12}\Tilde{\phi}^3.
\end{equation}
The resulting dynamical equation above is equivalent to the inflaton model in four-dimensional spacetime with potential \(\Tilde{V}=(\sigma/48)(\Tilde{\phi}^4-12\Tilde{\phi}^2)\). Furthermore, the dynamics of \(H\) is also can be solved from Einstein's equations evaluated on the visible brane, namely
\begin{eqnarray}\label{EinsteinOri1}
    3H^2&=&\alpha\left[\dot{\Tilde{\phi}}^2+\frac{\sigma}{24}\left(\Tilde{\phi}^4-12\Tilde{\phi}^2\right)\right],\\
    3\dot{H}+6H^2&=&\alpha\left[\dot{\Tilde{\phi}}^2-\frac{\sigma}{24}\left(\Tilde{\phi}^4-12\Tilde{\phi}^2\right)\right],
\end{eqnarray}
In conclusion, the inflaton model in four-dimensional space is the low energy limit of the fluctuating six-dimensional baby-Skyrme brane with a condition that the bulk cosmological constant must be positive. The constant \(\sigma\) is related to the bulk cosmological constant through \(\Lambda=2\alpha\sigma\) that is derived from the asymptotic behaviour of Einstein's equations and their corresponding boundary conditions. An interesting feature of the model shown above is that it is possible to have higher power terms in the potential by considering higher order perturbation on \(\xi\) and the vacuum expectation values (vev) of the inflaton are fully governed by the profile of the Skyrmion in the bulk.

In the general case where the dynamic of \(H\) cannot be neglected, we have the following set of dynamical equations
that are given in matrix form as follows
\begin{equation}\label{MatDyn1}
    \frac{d}{dt}\begin{bmatrix}\tilde{\phi}\\ p_{\Tilde{\phi}}\\ H\end{bmatrix}=\begin{bmatrix}
        p_{\Tilde{\phi}}\\
        -3Hp_{\Tilde{\phi}}+\frac{\sigma}{2}\Tilde{\phi}-\frac{\sigma}{12}\Tilde{\phi}^3\\
        -\alpha p_{\Tilde{\phi}}^2
    \end{bmatrix}
\end{equation}
where we have defined \(p_{\Tilde{\phi}}\equiv \dot{\Tilde{\phi}}\) and the system above is constrained by \eqref{EinsteinOri1}. The critical points \((\tilde{\phi}_c,p_{\tilde{\phi},c},H_c)\) are either \((0,0,H_c)\) or \((\pm\sqrt{6},0,H_c)\) where \(H_c\) depends solely on \(\tilde{\phi}_c\) through equation \eqref{EinsteinOri1}. Let \(x=(x_1,x_2,x_3)\) be a small perturbation around critical points such that \(\tilde{\phi}=\tilde{\phi}_c+x_1\), \(p_{\tilde{\phi}}=x_2\) and \(H=H_c+x_3\), where \(\tilde{\phi}_c\) and \(H_c\) are critical points of \(\tilde{\phi}\) and \(H\) respectively. We have the matrix equation \(dx/dt=Ax\) with
\begin{equation}
    A=\begin{bmatrix}
        0&1&0\\
        \frac{\sigma}{4}\left(2-\tilde{\phi}_c^2\right)&-3H_c&0\\
        0&0&0
    \end{bmatrix}.
\end{equation}
The eigenvalues of \(A\) can be written fully in \(\tilde{\phi}_c\), namely \(\{0,\lambda_{1,+},\lambda_{1,-}\}\) where
\begin{equation}
    \lambda_{1,\pm}=-\frac{1}{2}\left(3\sqrt{\frac{\alpha\sigma}{72}\left(\tilde{\phi}_c^4-12\tilde{\phi}_c^2\right)}\pm\sqrt{\sigma\left(8\alpha\tilde{\phi}_c^4-(1+96\alpha)\tilde{\phi}_c^2+2\right)}\right).
\end{equation}
We can see that for \(\tilde{\phi}_c=0\), one of the eigenvalues is positive, hence the critical point \((0,0,H_c)\) is a saddle point. On the other hand, the second critical point, \((\pm\sqrt{6},0,H_c)\), is also a saddle point if \(\alpha<-1/72\) but it becomes a stable point for the case of \(\alpha\geq -1/72\) since \(\lambda_{\pm}\) becomes purely imaginary for this case which implies that the solutions are oscillatory near this critical point. From the physical point of view, \(\alpha\) is non-negative, hence the second point is always a stable point. 

\begin{figure}
    \centering
    \includegraphics[width=0.8\textwidth]{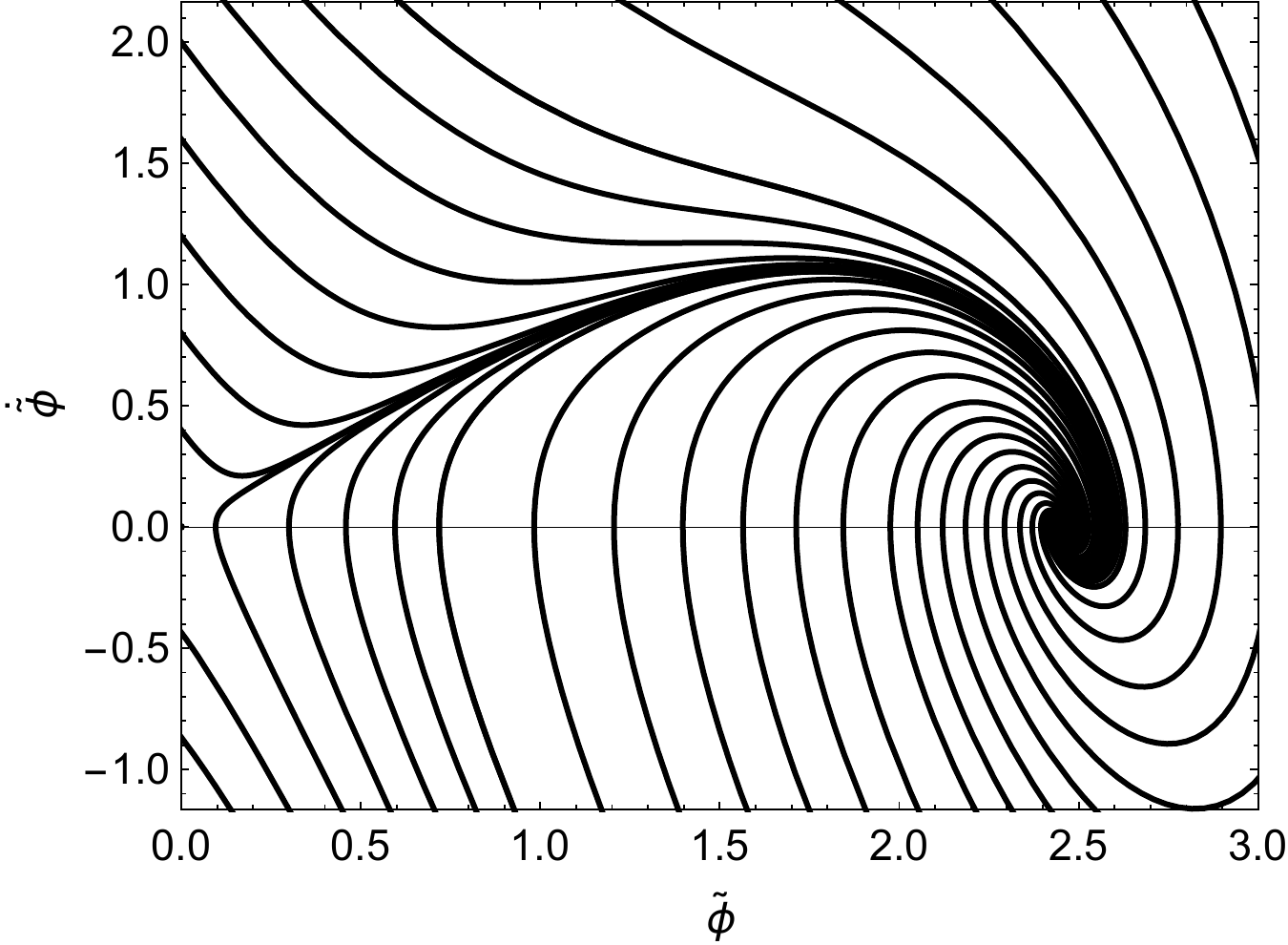}
    \caption{The trajectory of \(\Tilde{\phi}\) near unstable critical point, \(\Tilde{\phi}=0\), and stable point, \(\Tilde{\phi}=\sqrt{6}\), in phase space for several initial conditions with \(H=2/3\) and \(\sigma=5\).}
    \label{fig:PhaseSpaceInflaton}
\end{figure}

If we impose the condition for slow-roll inflation by assuming constant \(H\), then we can approximate \(\tilde{\phi}\) to be always near the stable equilibrium at \(\Tilde{\phi}=\sqrt{6}\). As such, the solutions of \(\Tilde{\phi}\) is approximately
\begin{equation}
    \Tilde{\phi}\approx \pm\sqrt{6}+\Phi_1 e^{-\frac{3H}{2}t}\sin\left(\sqrt{\sigma-\frac{9H^2}{4}}t+\Phi_2\right),
\end{equation}
where \(\Phi_1\) and \(\Phi_2\) depend on the initial condition. The corresponding example of trajectory in phase space for this dynamics with nearly constant \(H\) is given in Figure \ref{fig:PhaseSpaceInflaton}. Such decaying feature of \(\Tilde{\phi}\) is typical in the inflaton model and we are going to compare this result with the ones we get for the case of higher energy scale in the next section.
\section{High Energy Case}\label{general}
For the high-energy cases, the Skyrme term cannot be neglected anymore and it may contribute to the dynamics of the scalar field on the brane. Thus, in this section, we assume that in general \(\lambda\neq0 \) and \(k\neq0\) but we are still going to assume that \(\tilde{\phi}\) is a small perturbation around the static solutions \(\psi(z)\). 

Such higher energetic cases are relevant for both eras before the ordinary inflation and shortly after it ends, known as the reheating era. Here, we are going to show that the Skyrme term contributes to different decay behaviour of the oscillation of \(\Tilde{\phi}\) which depends on the value of \(\lambda\), and the reheating era dynamics with much faster decay of \(\Tilde{\phi}\) is reproduced even if interacting matter fields are not present.

\subsection{Localization On The Brane}
By following the similar procedure proposed in the previous section, we arrive at the following equations for the static unperturbed functions \(\psi\), \(F\), and \(G\),
    \begin{eqnarray}
    &&-\frac{3\left(F'+z
   F''\right)}{2zF}-\frac{\left(G'+z G''\right)}{2zG}+\frac{  G'^2}{2  G^2}-G\Lambda\\&=&\alpha \left[\left(\psi'{}^2+\frac{\sin^2\psi}{z^2}\right)+\lambda\left(\psi'{}^2\frac{\sin^2\psi}{Gz^2}\right)+\sigma G(1-\cos\psi)\right],\nonumber\\
    &&-\frac{
   F' G'}{FG}-\frac{3 F'^2}{2F^2}-\frac{2 F'}{zF}-G\Lambda\\&=&\alpha \left[\left(\psi'{}^2-\frac{\sin^2\psi}{z^2}\right)+\lambda\left(\psi'{}^2\frac{\sin^2\psi}{Gz^2}\right)-\sigma G(1-\cos\psi) \right]\nonumber,\\
    &&\frac{ F' G'}{F G}-\frac{ F'^2}{2
   F^2}-\frac{2  F''}{F}-G\Lambda\\&=&\alpha \left[\left(-\psi'{}^2+\frac{\sin^2\psi}{z^2}\right)+\lambda\left(\psi'{}^2\frac{\sin^2\psi}{Gz^2}\right)-\sigma G(1-\cos\psi)\right]\nonumber,\\
    &&\frac{2}{F^2Gz}\frac{\partial}{\partial z}\left(F^2z\psi'\left[1+\lambda \frac{\sin^2\psi}{Gz^2}\right]\right)\nonumber\\&=&\frac{2}{Gz^2}\sin\psi\cos\xi\left[1+\lambda\frac{\psi'{}^2}{G}\right]+\sigma\sin\psi,
\end{eqnarray}
where we have used the fact that \(n=1\) from the previous section.
By using the similar expansion near \(z=0\), given in \eqref{expandzero}, we have 
\begin{eqnarray}
    \psi(z)&=&0+\left(6G_0\alpha\sigma\right) z^2+O(z^3),\\
    F(z)&=&F_0-\left(\frac{F_0G_0\Lambda}{4}\right)z^2+O(z^3)\\
    G(z)&=&G_0+\left(\frac{G_0^2\Lambda}{4}\right)z^2+O(z^3).
\end{eqnarray}
We can see that the behaviour of all the functions near \(z=0\) is similar to the ones we found in the inflaton case which means that the brane with a lower energy scale is located at the asymptotic boundary. The study of localization within baby Skyrme brane for inflating visible brane in a more general manner has been done by \cite{Brihaye:2010nf} and it is found numerically that there exists a maximum cosmological constant where static solutions cease to exist beyond this value. 

From here, we have two shooting parameters, namely \(F_0\) and \(G_0\) that are going to be tuned in order to satisfy the asymptotic behaviour \eqref{expandinfty}. Through numerical simulations we found that \(G_0\) is approximately zero for any value of \(F_0\) and \(F_0\) depends only on the coupling constants \(\alpha\) \(\lambda\) and \(\sigma\). Since \(\lambda\) can be rescaled to one without any loss of generality by rescaling the coordinate, then we are left with only two free parameters for fine-tuning. Both of them depend on the physical parameters, namely the cosmological constant on the bulk, \(\Lambda\), and the bulk gravitational constant, \(\mathcal{G}\). The difference is that the gravitational constant on the bulk is directly related to the gravitational constant on the visible brane by \(G=\mathcal{G}F(z\rightarrow\infty)\), implying that the value is fixed by observation, but the cosmological constant on the bulk cannot be calculated directly from the observed cosmological constant on the visible brane. One of the possible ways to find the exact value of \(\Lambda\) is by finding the mass of inflaton that is directly related to the constant \(\sigma\) and the initial data of the inflation phase. For example, if the inflation starts near the vev of the \(\tilde{\phi}^4\) potential, then the mass is simply \(m=\sqrt{\sigma}\). This inflaton mass governs its oscillation frequency during the reheating era and the value should be related to the temperature of matter at the start of the hot big bang. 

%The value of \(F_0\) itself is useful to predict the energy scale at the brane located at \(z=0\). As mentioned in the previous paragraph that the value of \(F_0\) depends on both \(\alpha\) and \(\sigma\), we provide that behaviour of \(F_0\) as \(\alpha\) and \(\sigma\) varies in Figure \dots and Figure \dots. We can see that \dots
%
\subsection{Dynamic Fluctuation and The Reheating Era}\label{decay}
For this case, we cannot safely assume that \(\Tilde{\phi}\) is a global function on the extra dimensions. Thus we need to introduce the \(z-\) dependence in \(\Tilde{\phi}\), such that \(\xi(t,z)=\psi(z)+\tilde{\phi}(t,z)\). The dynamical equation for \(\tilde{\phi}\) near the visible brane is given by
\begin{eqnarray}\label{genTildePhi}
    0=\left(3H\dot{\tilde{\phi}}+\ddot{\tilde{\phi}}\right)\left[1+\lambda \Tilde{\phi}'{}^2\right]-\tilde{\phi}''\left[1-\lambda\dot{\Tilde{\phi}}^2\right]+4\lambda \dot{\Tilde{\phi}}\Tilde{\phi}'\dot{\tilde{\phi}}'-\frac{\sigma}{2}\Tilde{\phi}+\frac{\sigma}{12}\Tilde{\phi}^3+\dots,
\end{eqnarray}
and the Einstein's equations are
\begin{eqnarray}\label{EinsteinModif1}
    3H^2+\frac{3k}{a^2}&=&\alpha\left[\dot{\Tilde{\phi}}^2+\Tilde{\phi}'{}^2+\lambda\dot{\tilde{\phi}}^2\Tilde{\phi}'{}^2+\frac{\sigma}{24}\left(\Tilde{\phi}^4-12\Tilde{\phi}^2\right)\right],\\
    3\dot{H}+6H^2+\frac{3k}{a^2}&=&\alpha\left[\dot{\Tilde{\phi}}^2+\Tilde{\phi}'{}^2-\lambda\dot{\tilde{\phi}}^2\Tilde{\phi}'{}^2-\frac{\sigma}{24}\left(\Tilde{\phi}^4-12\Tilde{\phi}^2\right)\right].\label{EinsteinModif2}
\end{eqnarray}
We can see that variation of spatial curvature, \(k\), does not introduce any change in the dynamics of \(\Tilde{\phi}\).
We shall proceed by truncating the perturbation up to the third order, such that the stable equilibrium of the effective potential of for \(\Tilde{\phi}\) is located at \(\tilde{\phi}=\sqrt{6}\). 

In general, the dynamical system analysis of the system (\ref{genTildePhi}-\ref{EinsteinModif2}) is more complicated than the one we apply in the previous section due to the dependency on \(z\). The non-linear wave equation \(\eqref{genTildePhi}\) admits some families of solutions. It is known that the Skyrmions, \(\xi\), admits travelling wave solutions also known as solitons (\cite{Manton:2004tk,Fadhilla:2021jiz}), and we shall expect that \(\tilde{\phi}\), as the fluctuations of Skyrmion in this model, admits similar behaviour, at least as one of the possible families of solutions. Here, let us focus on the one-parameter family of solutions of \(\tilde{\phi}\) parameterized by wave number \(\bar{k}\). Each of the solutions in the family is an independent mode of \(\tilde{\phi}\) and the spatial dependence is separable from the time dependence, given in the form of the harmonic function \(e^{i\bar{k}z}\) such that each of the modes is eigenfunctions of \(\partial_z\). Thus, every spatial derivative satisfies \(\tilde{\phi}'=i\bar{k}\tilde{\phi}\). Substituting this relation for wave-like solutions to the general dynamical equations (\ref{genTildePhi}-\ref{EinsteinModif2}) and then followed by evaluating equation \eqref{genTildePhi} at the visible brane gives us the following matrix equation,
\begin{equation}\label{MatDyn2}
    \frac{d}{dt}\begin{bmatrix}\tilde{\phi}\\ p_{\Tilde{\phi}}\\ h\\ H\end{bmatrix}=\begin{bmatrix}
        p_{\Tilde{\phi}}\\
        -3Hp_{\Tilde{\phi}}+\left(\frac{\sigma-2\bar{k}^2}{2}\Tilde{\phi}-\frac{\sigma}{12}\Tilde{\phi}^3+5\lambda \bar{k}^2\phi p_{\Tilde{\phi}}^2\right)/\left(1-\lambda \bar{k}^2\tilde{\phi}^2\right)\\
        -2hH\\
        -\alpha p_{\Tilde{\phi}}^2\left(1-\lambda \bar{k}^2\tilde{\phi}^2\right)+h
    \end{bmatrix},
\end{equation}
where we have defined a new dynamical quantity \(h\equiv k/a^2\). It shall be understood that \(\tilde{\phi}\) in The system \eqref{MatDyn2} is only a single variable function, \(\tilde{\phi}(t,z\rightarrow\infty)\) and the system \eqref{MatDyn2} is constrained by \eqref{EinsteinModif1}. Similar to the case of low energy, here we have two types of critical points. The first one is \((0,0,0,H_c)\) and the second one is \((0,\pm\sqrt{6(1-2(\bar{k}^2/\sigma))},0,H_c)\) and \(H_c\) depends on \(\tilde{\phi}_c=0,\pm\sqrt{6(1-2(\bar{k}^2/\sigma))}\) through \eqref{EinsteinModif1}. The non-zero wave number shifts the location of the non-zero critical point where all the critical points coincide if \(\bar{k}^2\geq\sigma/2\). We can see that here we also have \(\bar{k}\) as a stability parameter, other than \(\alpha\) and \(\sigma\). Introducing \(y=(y_1,y_2,y_3,y_4)\) as a small perturbation around critical points, we have the matrix equation \(dy/dt=B y\), where \(B\) is given by
\begin{equation}
    B=\begin{bmatrix}
        0&1&0&0\\
        \frac{\sigma}{4}\left(2\left(1-(2\bar{k}^2/\sigma)\right)-\tilde{\phi}_c^2\right)&-3H_c&0&0\\
        0&0&-2H_c&0\\
        0&0&1&0
    \end{bmatrix}.
\end{equation}
The eigenvalues of \(B\) are \begin{equation}\left\{0,-2\sqrt{\alpha\sigma\left(\tilde{\phi}_c^4-12\tilde{\phi}_c^2\right)/72},\lambda_{2,+},\lambda_{2,-}\right\}\end{equation} with \(\lambda_{2,\pm}\) are given by the following expression
\begin{eqnarray}
    \lambda_{2,\pm}&=&-\frac{1}{2}\left(3\sqrt{\frac{\alpha\sigma}{72}\left(\tilde{\phi}_c^4-12\tilde{\phi}_c^2\right)}\right.\nonumber\\&&\left.\pm\sqrt{\sigma\left(8\alpha\tilde{\phi}_c^4-(1+96\alpha)\tilde{\phi}_c^2+2-4\bar{k}^2\right)}\right).
\end{eqnarray}
\(\tilde{\phi}_c=0\) is again a saddle point, but unlike the low energy case, in this high energy case, the stability of the non-zero critical point depends on both \(\sigma\) and \(\alpha\). This critical point becomes stable if \(\lambda_{2,\pm}\) is purely imaginary, which is equivalent to the following condition
\begin{equation}\label{stableCondi2}
    -\sigma^2+(3-\sigma)\sigma \bar{k}^2\leq 72\alpha(\sigma^2-4\bar{k}^4).
\end{equation}
The stability condition for the low energy case is recovered from \eqref{stableCondi2} at the extreme limit, \(\bar{k}=0\), hence \(\bar{k}\) must satisfy \(0<\bar{k}^2<\sigma/2\). Same with the low energy case, for the high energy case, \(\sigma\) is still a free parameter for the stability analysis, but the value can be found through physical arguments, as explained in section 4.1.

Now, if we introduce a small disturbance around the vev of \(\Tilde{\phi}\), \(\tilde{\phi}(t,z)=\sqrt{6}+\nu(t,z)\), then the dynamical field equation becomes
\begin{eqnarray}
    \left(3H\dot{\nu}+\ddot{\nu}\right)\left[1+\lambda \nu'{}^2\right]-\nu''\left[1-\lambda\dot{\nu}^2\right]+4\lambda \dot{\nu}\nu'\dot{\nu}'+\sigma\nu=0.
\end{eqnarray}
From the equation above, we can directly see the difference between the damping behaviour of low energy limit, which only came from the term proportional to \(\dot{\nu}\), and the more general case discussed here with Skyrme term turned on where we have additional damping term proportional to \(\dot{\nu}^2\). 

\begin{figure}
    \centering
    \includegraphics[width=0.45\textwidth]{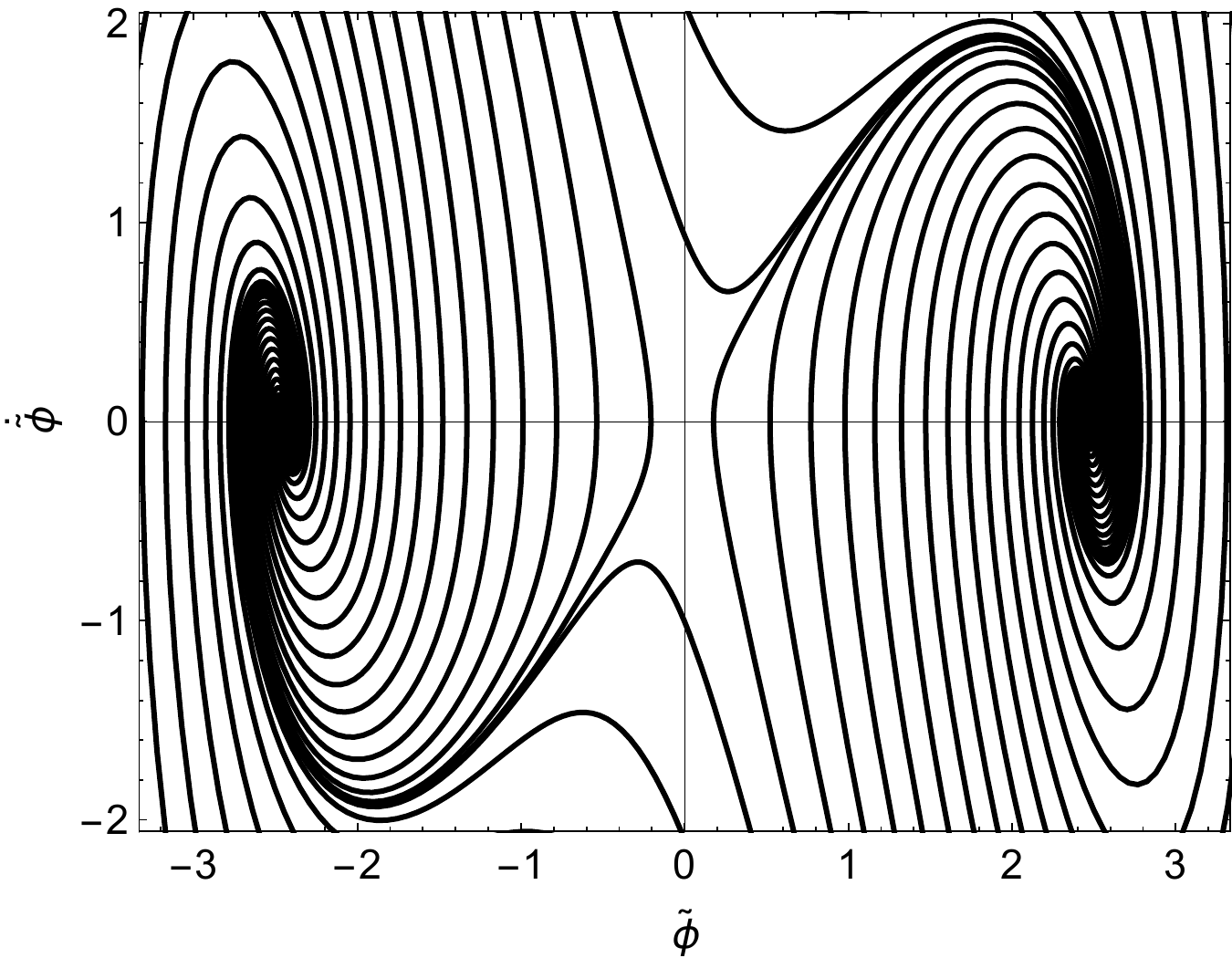}
    \includegraphics[width=0.45\textwidth]{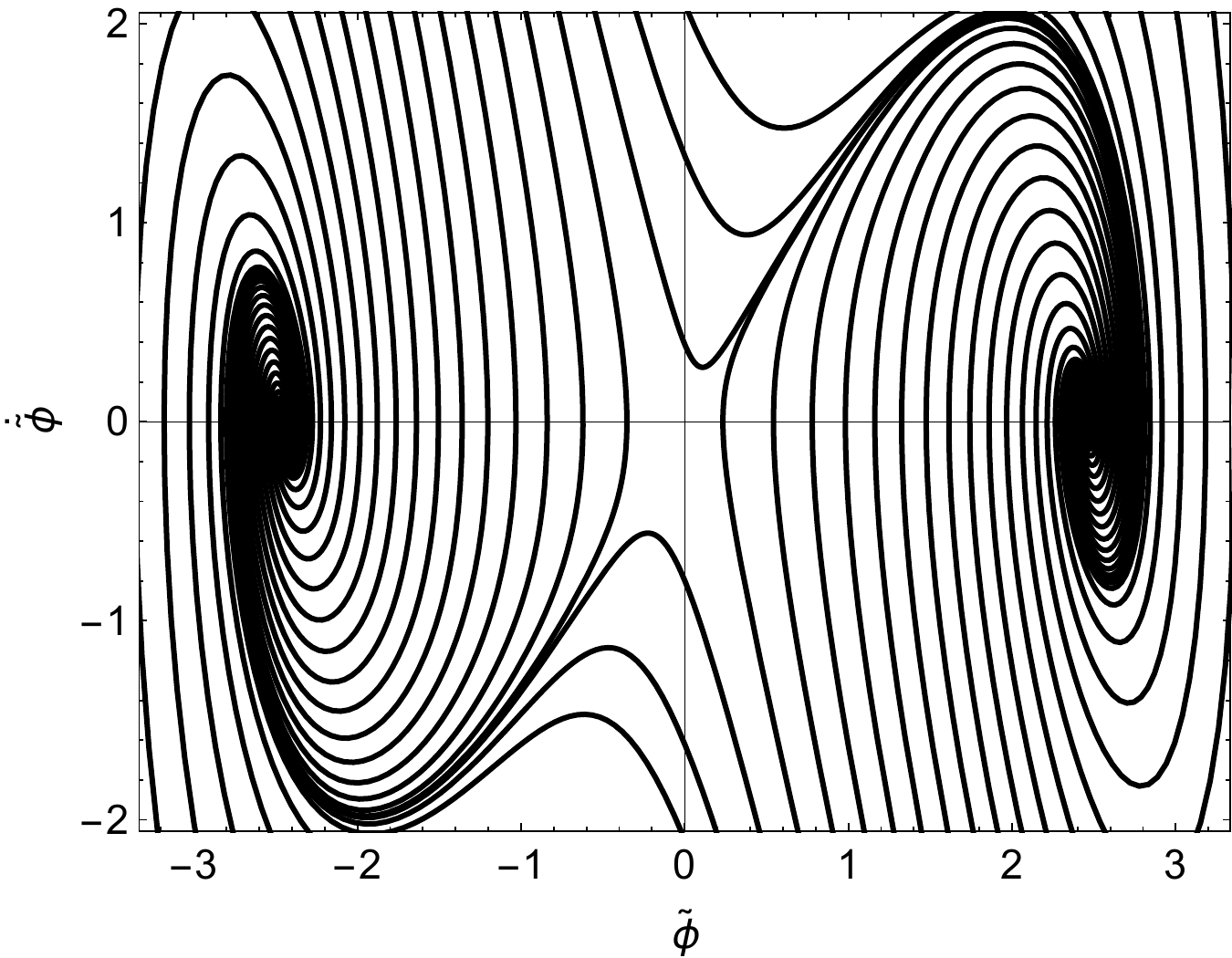}
    \caption{Comparison between the trajectories of \(\Tilde{\phi}\) with \(\lambda=1\) (left) and \(\lambda=5\) (right) in phase space.  Here we assume \(H=2/3\) and \(\sigma=10\).}
    \label{fig:PhaseSpacelambda13}
\end{figure}

\begin{figure}
    \centering
    \includegraphics[width=0.8\textwidth]{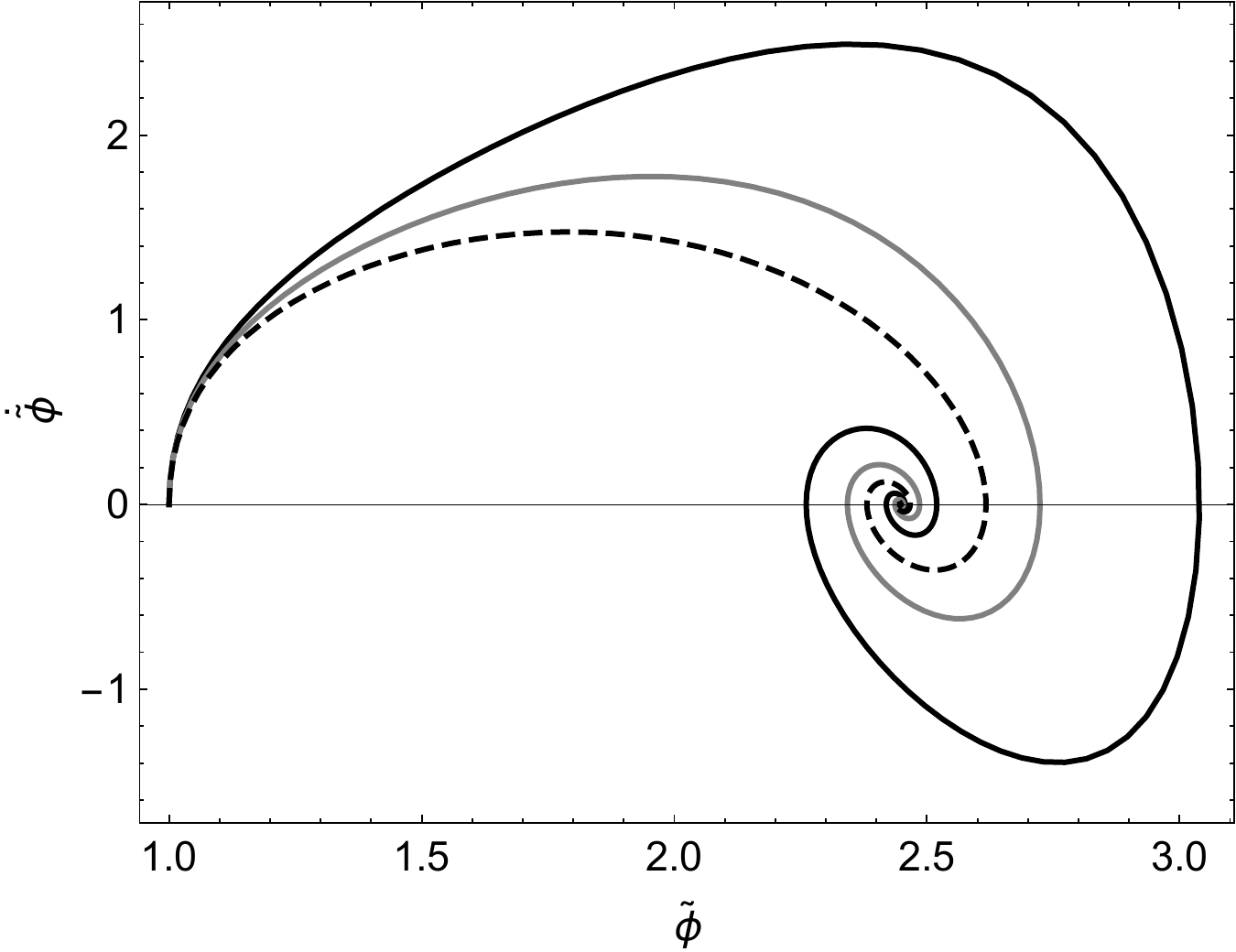}
    \caption{The trajectory of \(\Tilde{\phi}\) in phase space with initial condition \(\left(\tilde{\phi},\dot{\tilde{\phi}}\right)=(1,0)\) for \(\lambda=\) 4 (black), 0 (grey), and \(-4\) (dashed), with \(\sigma=10\) and \(H=3/2\) }
    \label{fig:Compare Lambda}
\end{figure}

In order to see the physical features of this equation, let us assume that \(H\) is approximately constant, each mode of the solution is a wave-like solution \(\nu\propto v_{\bar{k}}(t)e^{i\bar{k}z} \) and the leading modes with significant contribution possess small \(\bar{k}\),\footnote{This assumption implies that \(\nu\) is approximately independent of \(z\), that is the extreme limit we have at low energy, discussed in previous section} such that we can linearize the equation above which implies that the general solution is approximately superposition of the modes and factors \(e^{i\bar{k}z}\) in the dynamical equation can be approximated as one. As such, \(v_{\bar{k}}\) satisfies the dynamical equation
\begin{equation}
    \ddot{v}_{\bar{k}}+3H\dot{v}_{\bar{k}}-5\lambda \bar{k}^2v_{\bar{k}}\dot{v}_{\bar{k}}^2+\left[\sigma+\bar{k}^2\left(1+\lambda \sigma v_{\bar{k}}^2\right)\right]v_{\bar{k}}\approx 0.
\end{equation}
A particular solution that assumes well-behaviour of \(v_{\bar{k}}\) with respect to \(\bar{k}\) is given by the series \(v_{\bar{k}}=\sum_{i=0}^\infty a_i(t) \bar{k}^i\). This approach allows us to take only the leading terms of \(\bar{k}\) and neglect the terms with higher powers since \(\bar{k}\) is small. Substituting this solution to the differential equation gives us the following set of equations that needs to be solved for each coefficient of polynomials in \(\bar{k}\),
\begin{eqnarray}
    0&=&\ddot{a}_0+3H\dot{a}_0+\sigma a_0,\nonumber\\
    0&=&a_1,\nonumber\\
    0&=&\ddot{a}_2+3H\dot{a}_2+\sigma a_2+5\lambda a_0\dot{a}_0^2+a_0\left(1+\lambda a_0^2\right).
\end{eqnarray}
The equation for \(a_0\) is a damped oscillator equation with the solution given by
\begin{equation}
    a_0=\Phi_1 e^{-\frac{3H}{2}t}\sin\left(\sqrt{\sigma-\frac{9H^2}{4}}t+\Phi_2\right),
\end{equation}
where \(\Phi_1\) and \(\Phi_2\), again, depend on the initial condition. Assuming that the solution should be well-behaved under variation of \(\lambda\) and the inflaton limit (\(\lambda=0\)) is achievable, then we can approximate the solution of \(a_2\) up to leading order of \(\lambda\) as \(a_2= a_{2,0}+\lambda a_{2,1}\). This way, we can solve the contribution of the usual decay term proportional to \(\dot{a}_0\) and the quadratic decay term, \(\dot{a}_0^2\), separately. We are also going to assume that the oscillating factor is almost constant throughout the evolution of \(a_2\) because the decay is more dominant for this phase of inflation. This approximation gives us
\begin{eqnarray}
    a_2=-\frac{\Phi_1}{\omega^2}e^{-\frac{3H}{2}t}\sin\left(\omega t+\Phi_2\right)-\lambda\frac{\Phi_1^3}{\omega^2+9H^2}\left(\sigma+\frac{45H^2}{4}\right)\sin^3\left(\omega t+\Phi_2\right)
\end{eqnarray}
where \(\omega^2=|\sigma-(9H^2/4)|.\) The value of dominant \(k\) for the modes can be approximated by the mass-shell condition, \(\omega^2=m^2+\bar{k}^2\) at the bulk, which gives us \(\bar{k}\approx \pm(3H/2)\), implying that the spatial scale of the fluctuation on the extra dimensions is proportional to the expansion on the visible brane. As such, the solution of \(\tilde{\phi}\) near its stable equilibrium evaluated at \(z\rightarrow\infty\) is given by
\begin{eqnarray}
    \left.\Tilde{\phi}\right|_{z\rightarrow\infty}&\approx&\pm\sqrt{6}+ \Phi_1\left(1-\frac{9H^2}{4\omega^2}\right)e^{-\frac{3H}{2}t}\sin\left(\omega t+\Phi_2\right)\nonumber\\&&-\lambda\frac{9H^2\Phi_1^3}{4\left(\omega^2+9H^2\right)}\left(\sigma+\frac{45H^2}{4}\right)e^{-\frac{9H}{2}t}\sin^3\left(\omega t+\Phi_2\right).
\end{eqnarray}
From this result, we can find the effective decay time scale by first noticing that the solution of \(v_k\) is enveloped by a function \(p(t)\geq v_k(t)\) that is given by
\begin{equation}
    p(t)=\Phi_1\left|1-\frac{9H^2}{4\omega^2}\right|e^{-\frac{3H}{2}t}+\lambda\frac{9H^2\Phi_1^3}{4\left(\omega^2+9H^2\right)}\left(\sigma+\frac{45H^2}{4}\right)e^{-\frac{9H}{2}t}.
\end{equation}
The effective time scale can be found by integrating the normalized envelope, \(p(t)/p(0)\), from \(t=0\) to \(t\rightarrow\infty\). After some calculations we have
\begin{eqnarray}
    \tau&=&\left(\frac{2}{3H}\left|1-\frac{9H^2}{4\omega^2}\right|+\lambda\frac{H\Phi_1^2}{2\left(\omega^2+9H^2\right)}\left(\sigma+\frac{45H^2}{4}\right)\right)\left(\left|1-\frac{9H^2}{4\omega^2}\right|+\lambda\frac{9H^2\Phi_1^2}{4\left(\omega^2+9H^2\right)}\left(\sigma+\frac{45H^2}{4}\right)\right)^{-1}\nonumber\\
    &\approx& \frac{2}{3H}\left(1+\lambda\frac{H\Phi_1^2}{2\left(\omega^2+9H^2\right)}\left(\sigma+\frac{45H^2}{4}\right)\left|1-\frac{9H^2}{4\omega^2}\right|^{-1}\right),
\end{eqnarray}
where the assumption of small amplitude oscillation has been used. We can see that for \(\lambda=0\) we recover the result from the previous section and the time scale becomes longer for higher \(\lambda\). This implies that the term with quadratic \(\dot{\tilde{\phi}}\) could either slow down the decay, if \(\lambda>0\), or speed up the decay of the inflaton, if \(\lambda<0\).  We have also considered a more general approach where we numerically solve \(\Tilde{\phi}\) from equation \eqref{genTildePhi} and such behavior persists, even for initial conditions that are far from stable equilibrium, demonstrated in Figure \ref{fig:PhaseSpacelambda13}. It is also demonstrated numerically that the case with negative lambda will significantly accelerate the decay of \(\Tilde{\phi}\) in Figure \ref{fig:Compare Lambda}.
The faster (or, slower) decay of \(\Tilde{\phi}\) demonstrated above requires no interaction with any other fields, in contrast with the usual model where decay sped up due to the energy transfer to the matter fields. This behavior can be explained by the assumption of wave-like fluctuation in the bulk which implies that the energy might either get into or dissipate from the visible brane. Thus, this model allows energy transfer in and out of the brane.

The results above could explain the two different phases of inflation since the high-temperature environment could happen both before the standard inflation and at some points of the reheating phase. The case with a slower decay rate with \(\lambda>0\) might happen before the standard inflation which eventually coincides with the usual inflaton model when the universe is cold enough after rapid expansion. On the other hand, the reheating era started with the usual mechanism where inflaton transfers its energy density to the matter fields, then followed by a faster decay rate with \(\lambda<0\) when the universe is hot enough such that the Skyrme term cannot be neglected anymore. This approach implies that the coupling \(\lambda\) itself might be a dynamical quantity that varies slowly depending on the dynamics of the matter fields in the visible brane. This dependency on the matter field could explain the switching of signs before and after the standard inflation in this theory, but this statement requires more rigorous tests. A simple approach to theoretically test this statement would be considering a model of reheating with bosonic matter field, \(\chi\), confined in the visible brane and coupled to the Skyrme field through the coupling \(\lambda(\chi)\), and then followed by checking the energy scale of the universe at the end of the reheating phase where any loss of energy indicates a lower reheating temperature.

\section{Conclusion and Remarks}
We have shown in Section \ref{correspondence} that the small fluctuation on the Skyrme brane, represented by the time-dependent perturbative function \(\Tilde{\phi}\), can generate the inflaton model known for the four-dimensional spacetime that is the visible brane of our model with six-dimensional bulk. The standard inflation is recovered for the low energy limit where the sigma model term, quadratic to the derivative of the Skyrme field, dominates the Skyrme term that is quartic of the derivative of the Skyrme field. The resulting inflation model has a potential term that is not arbitrary and the expression of the potential term is fully governed by the Skyrmion evaluated at the brane. This model is shown to localize the gravity at the brane for a specific boundary condition and the lowest energy scale is located at the asymptotic boundary. We found that all relevant dynamics in the standard inflation with scalar field can be reproduced by this model and for a specific case where up to second-order perturbation is considered, the resulting model resembles inflation theory with \(\tilde{\phi}^4\) potential.

For the high-energy cases discussed in Section \ref{general} where the Skyrme term cannot be neglected, we found a different behavior of the dynamics of \(\tilde{\phi}\) where the field can either decay faster or slower, depending on the value of the coupling \(\lambda\). The faster decay case with \(\lambda<0\) is relevant to the reheating phase where conventionally the faster decay comes from the interaction between inflaton and the matter field, but in our model, the decay can still be sped up even when no matter field is present. We argue that this behavior is attributed to the wave-like property of the fluctuation which allows energy transfer in and out of the visible brane. Such property leads to a lower reheating temperature and can be possible evidence of the braneworld cosmology.

As mentioned earlier in subsection \ref{decay}, there is a possibility that \(\lambda\) changes sign throughout the inflation era, and one way to explain this possibility is by introducing a non-standard coupling \(\lambda(\chi)\) that depends on the matter field, \(\chi\). This coupling might also give us a picture of how the reheating process happens in contrast with the standard inflation. Another possible problem also came from the fact that the fluctuation is studied perturbatively which cannot give us the full picture of the gravity in the bulk. As such, the transfer of energy to the higher dimensional bulk cannot be fully understood. These problems are going to be addressed in future works. 
\section{Acknowledgement}The work in this paper is supported by GTA Research Group ITB and ITB Research Grant. E. S. F. would like to acknowledge the support from BRIN through the Research Assistant Programme 2023. B. E. G. would also acknowledge the support from BRIN Visiting Researcher Programme 2023.
\bibliography{main}
\bibliographystyle{unsrt}
\end{document}